\documentstyle[psfig]{mn}

\def\ltsima{$\; \buildrel < \over \sim \;$}
\def\lsim{\lower.5ex\hbox{\ltsima}}
\def\gtsima{$\; \buildrel > \over \sim \;$}
\def\gsim{\lower.5ex\hbox{\gtsima}}

\begin{document}

\title[Internal shocks in the jets of radio--loud quasars]
{Internal shocks in the jets of radio--loud quasars}

\author[Spada, Ghisellini, Lazzati, \& Celotti]
{Maddalena Spada$^{1}$, Gabriele Ghisellini$^2$,
Davide Lazzati$^{2,3}$
\& Annalisa Celotti$^4$\\
$^1$  Osservatorio Astrofisico di Arcetri, Largo E. Fermi 5, I--50125 Firenze\\
$^2$ Osservatorio Astronomico di Brera, Via Bianchi 46, I--23807 
Merate (Lc), Italy \\
$^3$ Present address: Institute of Astronomy, University of Cambridge, 
Madingley road, Cambridge CB3 0HA, U.K.\\
$^4$ S.I.S.S.A., Via Beirut 2/4, I--34014 Trieste, Italy\\
E--mail: {\tt spada@arcetri.astro.it, gabriele@merate.mi.astro.it, 
lazzati@ast.cam.ac.uk, celotti@sissa.it}}
\maketitle

\begin{abstract}
The central engine causing the production of jets in radio sources may
work intermittently, accelerating shells of plasma with different
mass, energy and velocity.  
Faster but later shells can then catch up slower earlier ones.  
In the resulting collisions shocks develop, converting some of the
ordered bulk kinetic energy into magnetic field and random energy of
the electrons which then radiate.  
We propose that this {\it internal shock scenario}, which is the
scenario generally thought to explain the observed gamma--ray burst
radiation, can work also for radio sources in general, and for blazar
in particular.  
We investigate in detail this idea, simulating the birth, propagation
and collision of shells, calculating the spectrum produced in each
collision, and summing the locally produced spectra from those regions
of the jet which are simultaneously active in the observer's frame.
We can thus construct snapshots of the overall spectral energy
distribution as well as time dependent spectra and light curves.  
This allows us to characterize the predicted variability at any
frequency, study correlations among the emission at different
frequencies, specify the contribution of each region of the jet to the
total emission, find correlations between flares at high energies and
the birth of superluminal radio knots and/or radio flares.  
The model has been applied to qualitatively reproduce the observed
properties of 3C 279. Global agreement in terms of both spectra and
temporal evolution is found.
In a forthcoming work, we explore the constraints which this scenario sets on 
the initial conditions of the plasma injected in the jet and the 
shock dissipation for different classes of blazars. 

\end{abstract}

\section{Introduction}
\label{uno}

Recent developments in the blazar field, mainly stimulated by the
discovery that these sources are strong $\gamma$--ray emitters, are
offering new and important clues for understanding relativistic jets.
All blazar spectral energy distributions (SED) are characterized by
two broad emission peaks e.g. \cite{F98}, strongly variable on
different timescales (Wagner \& Witzel 1995; Maraschi et al. 1994;
Hartman et al. 1996).  
The location of these peaks (their frequency,
relative and absolute flux) and their variability behavior are a
strong diagnostic tools to discriminate among theoretical models and
find the intrinsic physical parameters of the source.  
The broad band spectrum from the far infrared upwards can be explained
by a ``single-zone" model, where the two peaks are respectively due to
synchrotron and inverse Compton scattering from the same relativistic
electron population (Maraschi, Ghisellini \& Celotti 1992; Sikora,
Begelman \& Rees 1994).  
Since the radiative cooling time is much shorter than the dynamical one, 
the emitting particles must be accelerated {\it in situ} all along the 
jet, by means of a process not yet univocally determined.

The high energy flux varies generally on a timescale of the order 
of a day, and the bulk Lorentz factor (BLF) $\Gamma$ of the jet is 
estimated around $10$ \cite{G93}, leading to an upper limit for 
the source dimension of the order of $R\approx 10^{17}$ cm.
Furthermore, the required transparency for $\gamma$--rays to the
process of photon--photon pair production, sets a lower limit in the
most powerful sources, of order $\approx 10^{17}$ cm, to the typical 
distance where the bulk of the radiative dissipation occurs.

Blazars are also characterized by strong radio emission, with a flat
spectrum [$F(\nu)\propto \nu^{-\alpha}$, with $\alpha$ between $\sim
-0.5$ and $\sim 0.5$] and variability time scales of the order of
weeks, months (see e.g. Ulrich, Maraschi \& Urry, 1997).  
While the emission at frequencies from
the far infrared upwards can be consistently due to a single
population of electrons in the inner parts of the jet, the emission in
the radio band has to be generated -- in order not to be
self--absorbed -- on the outer, parsec scales where the
plasma properties are expected to be 
different. The bulk motion is still relativistic
at such distances, with BLFs similar to those inferred 
on the sub--parsec scales, $\Gamma\sim 10$ \cite{Z97}.

The existing models, both homogeneous (e.g. Sikora, Begelman \& Rees
1994; Inoue \& Takahara 1996) and inhomogeneous ones (e.g. Celotti,
Maraschi \& Treves 1991; Ghisellini \& Maraschi 1989), estimate the
jet plasma properties, the magnetic field $B$ and the random
Lorentz factors (RLF) of the electrons emitting at the peak of the 
spectra, $\gamma_{\rm p}$, by fitting the observed SED.
In the inner jet $B\approx 1$ G and $\gamma_{\rm p}$ varies from
$\sim$ 10$^2$ to 10$^6$ depending on the power of the source
\cite{G98}: for the powerful OVV quasars -- with bolometric luminosity
of the order of $L\approx 10^{47}-10^{48}$ erg s$^{-1}$ -- the
synchrotron peak is at infrared-optical frequencies and $\gamma_{\rm
p}$ is estimated around 10$^{2-3}$, while for the less powerful
sources, like the weak BL Lacs -- with $L\approx 10^{44}-10^{45}$erg
s$^{-1}$ -- the synchrotron emission peaks in the UV--X-ray range and
the RLFs of the electrons must be higher, $\gamma_{\rm p}\sim$ 10$^{4-6}$. 
This trend of decreasing peak energy with increasing source power
appears to be systematic and can be accounted for by a corresponding
increase in the intensity of a photon field external to the jet, which
enhances the inverse Compton cooling losses (Fossati et al. 1998;
Ghisellini et al. 1998).
In the parsec scale jet the observed (radio) emission is believed to
be synchrotron and the corresponding magnetic field and electrons
RLF are estimated to be $B\approx 10^{-3}$ G and
$\gamma_{\rm p}\approx 100$, leading to particle cooling timescales
much longer than the dynamical one.

Although these studies have allowed us to consistently derive the
physical parameters in the emitting region, many aspects in the
understanding of relativistic jets remain open, most notably
the jet energetics and particle acceleration.  In order to
explore these issues and their relationship we consider a scenario
where the plasma characteristics are not treated as free parameters,
but are the result of the jet dynamics, thus relating the observed
emission properties to the transport of energy along the jet.  We
achieve that by {\it quantitatively} considering a general scenario for the
interpretation of blazars analogous to the standard
scenario proposed to explain gamma--ray bursts, which was actually
originally proposed for blazars more than two decades ago (Rees 1978).

The most important assumption in this scenario is that the central
power engine produces energy which is channeled into jets in an
intermittent way, though such a time dependent process cannot be
easily inferred from first principles. 

If the physical conditions determining the energy deposition are not
steady, the distribution of BLFs and masses is
non-uniform within the ejecta: faster portions of the flow would then
catch up with slower ones leading to the development of relativistic
shocks.  The shocks would plausibly heat the expanding ejecta,
generate/amplify a tangled magnetic field and accelerate leptons to
relativistic energies, in turn causing synchrotron and inverse Compton
emission.  In this scenario -- named internal shock model (e.g.  Rees
\& Meszaros 1994; Lazzati, Ghisellini \& Celotti 1999, Panaitescu,
Spada \& Meszaros 1999) -- the following observational characteristics 
can be naturally explained:

\begin{itemize}
\item{\bf Efficiency --} 
The radiative output of radio sources in
general, and blazars in particular, has to be a small fraction of the
energy transported by the jet (less than 10\%): extended radio
structures require in fact a power input exceeding what is emitted
(e.g. Rawlings \& Saunders 1991; Celotti \& Fabian 1993).  
Low efficiency is indeed a characteristic of the internal shock model,
as the range of the BLFs of the colliding shells can
be only small (say the ratio of the maximum and minimum ones
$\Gamma_{\rm M}/\Gamma_{\rm m}\sim 3$) in order to have a
resulting jet bulk motion on larger scale with $\Gamma\approx 10$.
\item {\bf Minimum distance for dissipation --} 
As mentioned in the Introduction, there is a minimum dimension for the
$\gamma$-ray source and the bulk of the observed power must be
produced at some distance from the jet apex, i.e. from the accretion
disk.  In other words, the emitting zone must be far away from
$X$-rays sources and not extremely compact itself otherwise, quite
inevitably, the high energy $\gamma$--rays are absorbed by the dense
radiation fields. The electron--positron pairs generated by such
photon--photon interaction would then re-process the $\gamma$--ray
power into softer (especially X--ray) radiation, which is not observed
(Ghisellini \& Madau 1996).
In the internal shock model the jet becomes radiative on scales much
larger than the central source dimension ($R_{0}\approx 10^{14}$cm for
a black hole of $M=10^{9}M_{\odot}$) since all the velocities are
close to $c$ and therefore the shells propagate for a significant
distance before collisions can start.  The expected minimum distance
for the $\gamma$-rays emission is of the order of $R\approx
R_{0}\Gamma^2\approx 10^{16}-10^{17}$cm, where $R_0/c$ measures the
time interval between two consecutive shell ejections.
\item {\bf Variability --} 
Clearly the observed large amplitude variability must be explained
by a non steady--state model: we believe that the shell--shell
collision scenario studied here is the simplest among them.
\item {\bf Continuous energy deposition along the jet --}
Dissipation is observed to occur along the whole jet, from the inner
to the parsec scales with decreasing luminosity and typical photon
frequency. The dynamical evolution of a wind in the internal shock
model can indeed reproduce such behavior: shell--shell collisions
take place along the whole jet, leading to a continuous conversion of
bulk energy into internal one over a large range of distances.
Due to the averaging of the shell BLFs during the
expansion, the collisions become progressively less efficient as the
distance increases, thus progressively reducing the released internal
energy, the electron RLF, and the magnetic field.
\end{itemize}

As this scenario allows to predict a general and fairly complete
description of the {\it entire} radiative jet, we believe that it can
be a powerful tool to understand blazar spectra, their variability and
the possible correlations between observed quantities on different
scales.  While we plan to fully explore the large potentialities of
this model in a forthcoming work, in this paper we present the basic
assumptions and the general set up of the model.  For illustration
purposes (i.e. not aiming to determine precise values of the physical
quantities) we also consider the case of a specific well studied
blazar, specifically 3C 279.  Section 2 therefore describes the
dynamics of the wind and the shocks forming from shell collisions,
while in Section 3 the radiative properties of the shocked plasma are
presented. The numerical simulations for the time dependent evolution
of the system are described in Section 4, while the results and the
specific case of 3C 279 are the subject of Section 5. In the final
Section 6 we discuss our findings.

\section{Hierarchical internal shocks}
\label{sec:para}

The entire jet emission is simulated by adding pulses radiated in a
series of internal shocks that occur in a unstable relativistic wind
(Daigne \& Mochkovitch 1998; Spada, Panaitescu \& Meszaros 2000).
After (parametrically) setting the dynamics of the wind ejection, we
calculate the radii where the shells collide.  
For each collision we study the hydrodynamics, in order to determine
the shock velocity, the compression ratio and the internal energy
$E_{\rm sh}$ of the shocked fluid.  Assuming a given partition of $E_{\rm sh}$
among protons, electrons and magnetic field we calculate the relevant
physical parameters in the shocked fluid, and finally the features of
the emitted radiation.

\subsection{Ejection features}

The wind is discretized as a sequence of $N=t_{\rm w}/t_{\rm v}$
shells, where $t_{\rm w}$ is the duration time of the wind ejection from the
central source and $t_{\rm v}\ll t_{\rm w}$ is the average interval
between consecutive ejections. Each shell is characterized by a mass
$M_j$, a bulk Lorenz factor $\Gamma_j$ and an ejection
time $t_j$: at $t=t_j$ the j-th shell is at a distance
$R_j=R_0$, where $R_0$ is the central source dimension
($R_0=10^{14} $cm).

The shell BLFs $\Gamma_j$ are chosen randomly from a uniform distribution
between $\Gamma_{\rm m}$ and $\Gamma_{\rm M}$, and the shell masses are
also randomly picked up from a uniform distribution with an average
value $M_{\rm t}/N$, where the total mass in the wind $M_{\rm t}$ 
is determined by the requirement:
\begin{equation}
\sum_{\rm j=1}^N M_j\Gamma_j c^2= L_{\rm w} t_{\rm w}
\end{equation}
where $L_{\rm w}$ is the wind luminosity. 
The time interval between the ejection of a couple of consecutive
shells is assumed to mimic a scenario in which the inner engine
behaves as an accumulation mechanism: the quiescent time between the 
ejection of the $j^{th}$ and the $(j+1)^{th}$ shells is proportional 
to the total energy in the $(j+1)^{th}$ shell.

The ejected shell width $\Delta_j$ is of the same order of the
central source dimension $R_0$ and the shell internal energy $U_j$ is
negligible (all the energy is stored as bulk kinetic energy).

\subsection{Wind dynamics}\label{WD}

The dynamics of the wind expansion is characterized by a series of
two-shell collisions in which the faster shells catch up with the
slower ones in the outer parts of the ejecta.
The calculation of the collision radii follows the procedure of
Kobayashi, Piran \& Sari (1998). 
Given the wind configuration after the $i$-th collision, i.e.  
shell radius $R^i(j)$, shell velocity $\beta^i(j)$, shell width 
$\Delta^i(j)$ and expansion velocity of the shell $\beta_{\rm e}^i(j)$, 
we calculate the minimum value among all the collision times
between pairs of shells ($j, j-1$) with $\beta_{j-1}>\beta_j$.
\begin{equation}
\delta t_{i+1}={\rm min}(t_{j,j-1})
\end{equation}
\begin{equation}
t_{j,j-1}=\frac{R(j)-R(j-1)-0.5(\Delta(j)+\Delta(j-1))}
{\beta(j-1)-\beta(j)+0.5(\beta_{\rm e}(j)+\beta_{\rm e}(j-1))}.
\end{equation}
The radius of the $(i+1)$-th collision represents the position of the two
shells colliding after $\delta t_{i+1}$. This procedure is iterated for 
a wind described by an up-dated configuration, with a merged shell 
instead of the two colliding ones.

Regarding the two-shell interaction as an inelastic collision, the 
momentum and energy conservation laws yield the physical properties 
of the merged shell.
If ($\Gamma_i$, $M_i$, $\eta_i$) and 
($\Gamma_{\rm o}$, $M_{\rm o}$, $\eta_{\rm o}$) are the
BLFs, rest masses and internal energies of the fast (inner) and 
slow (outer) shell with $\Gamma_i>\Gamma_{\rm o}$, the merged shell 
has a rest mass $M_{\rm m}=M_i+M_{\rm o}$, a BLF given by:
\begin{equation}
\Gamma_{\rm m}=\left(\frac{\mu_i\Gamma_i+\mu_{\rm o}\Gamma_{\rm o}}
{\mu_i/\Gamma_i+\mu_{\rm o}/\Gamma_{\rm o}}\right)^{1/2}
\label{eq:gfin}
\end{equation}
where $\mu_i = M_i+\eta_i/c^2$ and 
$\mu_{\rm o} = M_{\rm o}+ \eta_{\rm o}/c^2$.
The total internal energy of the merged shell will then be:
\begin{equation}
E_{\rm in}= \eta_i + \eta_{\rm o} + \mu_i c^2 
(\Gamma_i-\Gamma_{\rm m})+\mu_{\rm o} 
c^2(\Gamma_{\rm o}-\Gamma_{\rm m}).
\label{eq:ein}
\end{equation}

The dynamical efficiency, namely the newly generated internal energy
over the total energy, is given by:
\begin{equation}
\epsilon_{\rm d}={{\mu_i\,(\Gamma_i-\Gamma_{\rm m}) +
\mu_{\rm o}\,(\Gamma_{\rm o}-\Gamma_{\rm m})} \over
{\Gamma_i\,\mu_i+\Gamma_{\rm o}\,\mu_{\rm o}}}.
\end{equation}
This efficiency decreases with the ratio $\Gamma_i/\Gamma_{\rm o}$ and 
is maximized for $\mu_{\rm o}=\mu_i$.

The width of the shells spreads between two consecutive
collisions with a velocity:
\begin{equation}
\beta_{\rm e}=\frac{2 \beta_{\rm s}'}{\Gamma^2}
\frac{1}{1-(\beta\beta_{\rm s}')^2},
\end{equation}
as measured in the observer's frame,
where $\beta_{\rm s}' c = v_{\rm s}'$ is the sound velocity of the plasma
in the shell co--moving frame:
\begin{equation}
v_{\rm s}'=\sqrt{\frac{1}{3}\frac{E_{\rm th}'}{M_{\rm sh}}}.
\end{equation}
During the free expansion between two subsequent collisions,
each shell loses internal energy for adiabatic losses, while the 
BLF is held constant. 
The magnetic field, on the contrary, is generated
at each collision without keeping memory of the field
generated in precedent collisions (if any).
The plasma thermal energy $E_{\rm th}$ increases in each collision --
by the fraction of $E_{\rm in}$ which is not used to either accelerate
electrons to a power--law energy distribution and/or generate/amplify
magnetic field -- leading to an increase in the plasma sound velocity
$v_{\rm s}'$ during the wind expansion.

Since during each collision the propagation of the shocks through the
two shells compresses the width of the merged shell $\Delta_{\rm m}$,
the calculation of $\Delta_{\rm m}$ itself requires a hydrodynamic
treatment of the interaction, that is presented in the next section.

\subsection{Hydrodynamic of the two--shell interaction}

\begin{figure}
\psfig{figure=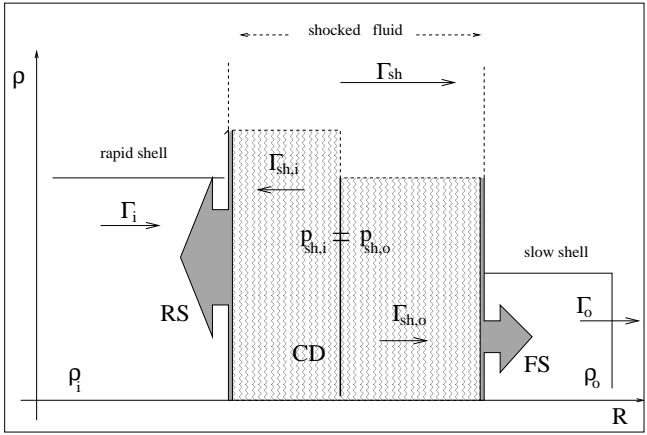,width=8.5cm,angle=-90}
\caption{\footnotesize{Diagram of a two-shell collision. A
inner fast shell with BLF $\Gamma_i$ and density $\rho_i$ catches
up with a slower shell characterized by 
BLF $\Gamma_{\rm o}$ and density $\rho_{\rm o}$.
The reverse shock (RS), propagating in the inner shell, and the
forward shock (FS), propagating in the outer shell, are shown.
Both of the shocked fluids move with BLF $\Gamma_{\rm sh}$
in the lab frame. In the un-shocked fluid frame the RS BLF
is $\Gamma_{{\rm sh},i}$ and the FS BLF is
$\Gamma_{\rm sh,o}$.
The co--moving pressures of the shocked fluids are equal around 
the contact discontinuity, $(p_{\rm sh})_i=(p_{\rm sh})_{\rm o}$.}}
\label{fig:shock}
\end{figure}

Let us consider the dynamics of the interaction between two shells.
Here it will be assumed that the presence any magnetic field is
negligible from the dynamics point of view.
During the collision two shocks are formed, separated by a contact
discontinuity: a forward shock (FS) that propagates in the external
slow shell and a reverse shock (RS) that propagates backward in the
faster shell (see Figure \ref{fig:shock}).

The calculation of the shocked plasma BLF, $\Gamma_{\rm sh}$,
in the lab frame can be done applying the following conditions \cite{PM99}:
\begin{itemize}
\item The Blandford--McKee (1976) jump equations for the FS and RS. 
For the FS the energy and particle number
conservation yield:
\begin{equation}
\left(\frac{e_{\rm sh}}{n_{\rm sh}}\right)_{\rm o}=m_{\rm p} 
c^2(\Gamma_{{\rm sh},o}-1)
\label{bmk1}
\end{equation}
\begin{equation}
\left(\frac{n_{\rm sh}}{n}\right)_{\rm o}=
\frac{\widehat{\gamma}\Gamma_{{\rm sh},o}+1}
{\widehat{\gamma}-1}, 
\label{bmk2}
\end{equation}
where $e_{\rm sh}$ and $n_{\rm sh}$ are the internal energy and the
numerical baryon density of the shocked plasma, respectively, and $n$
is the pre--shocked numerical density.  All these quantities are
measured in the fluid co-moving frame.  $\widehat{\gamma}$ is the
adiabatic index and $\Gamma_{\rm sh,o}$ is the BLF of the
shocked plasma in the frame of the outer shell.  A similar set of
equations holds for the RS.  The shocked plasma pressure
$p=(\widehat{\gamma}-1)u$ is calculated through Eqs. (\ref{bmk1}) and
(\ref{bmk2}): $(p_{\rm sh})_{o(i)}=(\Gamma_{\rm sh,o(i)}-1)
(\widehat{\gamma} \Gamma_{\rm sh,o(i)}+1)\rho_{o(i)}c^2$, with
$\rho_{o(i)}$ being the co--moving matter density of the shells before
the collision.
\item The condition of equal pressure for the shocked fluids around
the contact discontinuity (CD) $(p_{\rm sh})_{\rm o}=(p_{\rm sh})_i$,
and the condition of equal velocity around the CD yield a quartic
equation for the lab--frame BLF $\Gamma_{\rm sh}$:
\begin{eqnarray}
& 
\widehat{\gamma}(g^4-Y)x^4+
2(\widehat{\gamma}-1)g(Y-g^2)x^3+
2(2-\widehat{\gamma})g^2(Y-1)x^2 &
\nonumber \\
 &
+2(\widehat{\gamma}-1)g(g^2Y-1)x+
\widehat{\gamma}(1-g^4Y)=0,&
\label{eq}
\end{eqnarray}
where:
\begin{equation}
x=\frac{\Gamma_{\rm sh}}{\sqrt{\Gamma_{\rm o}\Gamma_i}}~,~~~~
g=\sqrt{\frac{\Gamma_i}{\Gamma_{\rm o}}}~~~~~~ \left(\frac{1}{g}<x<g\right),
\end{equation}
where $\Gamma_{\rm o}$ and $\Gamma_i$ are the initial BLFs of
the two shells and $Y$ is the ratio of the pre--shocks densities
of the slow and fast shells.
Eq.(\ref{eq}) is solved numerically.
In the range of interest and for any physical choice of
parameters Eq.(\ref{eq}) yields only one solution.
\end{itemize}

Once $\Gamma_{\rm sh}$ is known we can evaluate: 
\begin{itemize}
\item The shell widths after the shocks.
The compression ratio $\rho/\rho_{\rm sh}$ is
obtained by the jump conditions (\ref{bmk1}) and (\ref{bmk2}) (assuming
a value for the adiabatic index $\widehat\gamma=4/3$) and: 
\begin{equation}
\Delta_{sh,i}=\Delta_i\left(\frac{\rho_i}{\rho_{sh,i}}\right)_{\rm lab}
~~
\Delta_{sh,o}=\Delta_{\rm o}\left(\frac{\rho_{\rm o}}
{\rho_{sh,o}}\right)_{\rm lab}, 
\end{equation}
where $\Delta_i$ and $\Delta_{\rm o}$ are the pre-shock values.
\item The internal energies $E_{\rm RS}$ and $E_{\rm FS}$ for the RS and FS.
The equality of the two shocks velocities and of the comoving internal 
energy densities yields (in the lab frame) to:
\begin{equation}
\frac{E_{\rm RS}}{E_{\rm FS}}=\frac{\Delta_{{\rm sh},i}}{\Delta_{\rm
sh,o}}={\it R}.
\end{equation}
Thus the energy in the RS is $E_{\rm RS}={\it R}/(1+{\it
R})\times E_{\rm in}$, and in the FS is $E_{\rm
FS}={1}/(1+{\it R})\times E_{\rm in}$, where $E_{\rm in}$ is the total
internal energy generated during the collision (eq.
\ref{eq:ein}).
\item Knowing the shell width and the shock velocity we can then
estimate the shell crossing time of the shock $\delta t_{\rm
cr}=\Delta/|v_{\rm sh}-v_0|$ where $v_{\rm sh}$ and $v_0$ are the
post-shock and pre-shock material velocity, respectively.
\end{itemize}

\subsection{Shocked fluid parameters}

The parameters determining the synchrotron and inverse Compton
emission are the magnetic field strength $B$ and the electron energy
distribution in the shocked fluid.

Given the uncertainty in determining these quantities directly from
the physics of relativistic shocks, these are parametrized by
dimensionless parameters defined as: $\epsilon_{\rm B}$ and
$\epsilon_{\rm e}$, which measure the fraction of the co-moving
internal energy density $U'_{\rm sh}$ stored in magnetic field $U_{\rm
B}'$ and in electrons $U_{\rm e}'$, respectively (the apex indicates
quantities measured in the co-moving frame).

The co-moving magnetic field is then defined by 
$B'^2/(8\pi)=\epsilon_{\rm B} U'_{\rm sh}$:
\begin{equation}
B'=
\frac{1}{\Gamma_{\rm sh}}\sqrt{\frac{2\epsilon_{\rm B} E_{\rm sh}}{R^2\Delta}},
\end{equation}
and is assumed to be randomly oriented in space.

Following the results on non-relativistic shocks, and the recent
results on relativistic ones (Bednarz \& Ostrowski 1998; Kirk et al. 2000),
we also postulate that the electrons are accelerated (instantaneously)
behind the shock as a power law $N(\gamma)\propto\gamma^{-p}$ with
$\gamma>\gamma_{\rm b}$, and $p>2$.  
For an equal number of protons and 
electrons, the number density of non--thermal electrons is: $n'_{\rm
e}=\zeta_{\rm e}\rho_{\rm sh}/m_{\rm p}$, where $\zeta_{\rm e}$ is the
fraction of electrons effectively accelerated. Since $U_{\rm
e}'=\epsilon_{\rm e} U_{\rm sh}$, the minimum RLF
$\gamma_{\rm b}$ corresponds to:
\begin{equation}
\gamma_{\rm b}=\frac{p-2}{p-1}\left(1+\frac{\epsilon_{\rm e}}{\zeta_{\rm e}}
\frac{U_{\rm sh}'}{\rho_{\rm sh}'}\frac{m_{\rm p}}{m_{\rm e}}\right)\approx
1837\frac{p-2}{p-1}\frac{\epsilon_{\rm e}}{\zeta_{\rm e}}\Gamma_{\rm sh}.
\label{game}
\end{equation}

In conclusion, by studying the wind dynamics and shock hydrodynamics
we can determine for each collision the energy density in non--thermal
electrons and in magnetic field and estimate the minimum RLFs
$\gamma_{\rm b}$ of the electron energy distribution injected by the
shocks.

\section{The spectrum}
\label{sec:spec}

Having estimated such quantities, we can therefore calculate the
spectrum of the radiation emitted during each collision. In order to
achieve that we adopt some simplifications:

\begin{itemize}
\item The emitting zone is assumed to be homogeneous, and the plasma
is embedded in a tangled magnetic field.  Its comoving volume is
$V=\pi \psi^2 R^2\Delta R^\prime$, where $\psi$ is the half-opening
angle of the jet, assumed conical.

\item The relativistic particles are treated as having the same energy
distribution throughout the entire emitting region.  In a more
realistic approach, the particles would be accelerated at the shock
front, where most of the energetic electrons would be located, while
cooled particles would be progressively more distant from the shock
front.  However, as we are interested in the ``average" spectrum, we
assume to collect radiation emitted by the entire shell.  Clearly such
simplification does not allow to consider changes in spectral details
occurring on a timescale faster than the light crossing time of a
single shell.

\item There is a source of soft photons external to the jet.  We
identify it with the emission reprocessed in the broad line region
(BLR).  Therefore we consider a luminosity $L_{\rm ext}=a L_{\rm
disk}$ is produced within $R_{\rm BLR}$, corresponding to a
radiation energy density (as measured in the frame comoving with the
shell) given by (e.g. Ghisellini \& Madau 1996):
\begin{equation}
U_{\rm ext} \, =\, {17\over 12} \, { aL_{\rm disk} \Gamma^2 \over 4 \pi
R_{\rm BLR}^2 c}
\end{equation}
For simplicity, this seed photon component is considered 
to abruptly vanish beyond $R_{\rm BLR}$.

\item We assume that the particle distribution results from
the continuous injection and cooling processes, where particles 
are injected with a power law 
energy distribution of index $s=(p-1)$ between $\gamma_{\rm b}$ 
and $\gamma_{\rm max}$ (as measured in the comoving frame).
The corresponding power injected in the form of relativistic particles
is $L_{\rm e}$.
The equilibrium particle distribution $N(\gamma)$  is then found through
the following procedure.

\noindent 
We calculate the (comoving) random RLF $\gamma_{\rm cool}$
of those particles with a cooling time comparable with the shock
crossing time $t_{\rm cross}$ and above $\gamma_{\rm cool}$ the
distribution is taken as a power law of index $p$.  Below
$\gamma_{\rm cool}$ there are two cases:

i) if $\gamma_{\rm cool} > \gamma_{\rm b}$, we assume that $N(\gamma)
\propto \gamma^{-s}$ between $\gamma_{\rm b}$ and $\gamma_{\rm cool}$,
and $N(\gamma)=0$ for $\gamma < \gamma_{\rm b}$;

ii) if $\gamma_{\rm cool} < \gamma_{\rm b}$, $N(\gamma) \propto
\gamma^{-2}$ between $\gamma_{\rm cool}$ and $\gamma_{\rm b}$, and
$N(\gamma)=0$ for $\gamma <\gamma_{\rm cool}$.

The normalization of $N(\gamma)$ is determined according to 
whether particles with $\gamma\sim \gamma_{\rm b}$ can or cannot cool 
in the timescale $t_{\rm cross}$:

a) if most of the power injected in relativistic particles is radiated
in a short timescale ($<t_{\rm cross}$, this is equivalent to
demand $\gamma_{\rm b} > \gamma_{\rm cool}$) we apply a
luminosity balance condition:
\begin{equation}
L_{\rm e} \, =\, V m_{\rm e}c^2 \int \dot\gamma N(\gamma) d \gamma,
\end{equation}
where $V$ is the emitting volume and $\dot
\gamma$ is the total radiative cooling rate, which includes
synchrotron, synchrotron self--Compton (SSC) and external Compton (EC) losses. 
In order to 
allow for the possibility of having multiple Compton scatterings,
especially when $\gamma_{\rm b}$ is not very large, we
first calculate how many scattering orders $n_{\rm IC}$ can take place 
before the Klein--Nishina limit is reached (we consider up to 10
orders of scatterings). 
Taking $\gamma_{\rm b}$ as the relevant energy, the number of
scatering orders in the Thomson regime are 
\begin{equation}
n_{\rm IC}\, =\, {\ln (\gamma_{\rm b}/x_{\rm B}) \over 
\ln (4\gamma_{\rm b}^2/3)} \, -1
\end{equation}
where $x_{\rm B}\equiv h\nu_{\rm B}/(m_{\rm e}c^2)$ and $\nu_{\rm B}
=eB/(2\pi m_{\rm e} c)$ is the (non--relativistic) Larmor frequency.
We define  $U_{\rm e} \equiv L_{\rm e}/(\pi R^2 c)$ and introduce the
Comptonization parameter $y$ as:
\begin{equation}
y\, \equiv \, {3\over 4} 
\sigma_{\rm T} \Delta R^\prime \int \gamma^2 N(\gamma) d \gamma.
\end{equation}
Here $\Delta R^\prime$ is the thickness of the shell at the
time of the collision. For semplicitly, we neglect its 
further expansion while it is radiating.
With this definition of $y$, Eq. (18) can be re--written as:
\begin{equation}
U_{\rm e}\, = y U_{\rm B} \left( 1+{U_{\rm ext} \over U_{\rm B}} + y +
y^2 + \, ....\, y^{n_{\rm IC}}\right).
\end{equation}
b) if most of the power is not radiated in $t_{\rm cross}$, the electrons
retain their energy and
\begin{equation}
E_{\rm e} \, =\, L_{\rm e} t_{\rm cross}\, = \, m_{\rm e}c^2 V \int
N(\gamma)\gamma d\gamma.
\end{equation}
Assuming the same broken power--law shape for $N(\gamma)$ we can then
determine its normalization.

\end{itemize}

\subsection{Synchrotron emission}

To calculate the synchrotron emission we adopt 
the pitch angle averaged synchrotron emissivity for a single electron 
given in, e.g., Ghisellini \& Svensson (1991):
\begin{eqnarray}
& &j(\gamma, \nu)\, =\,
{3\sqrt{3} \over \pi} \, {\sigma_{\rm T} c 
U_{\rm B}\over \nu_{\rm B}}\, t^2 \times 
\nonumber \\
&\, &
\left\{K_{4/3}(t) K_{1/3}(t) - {3 t\over 5}
\left[ K^2_{4/3}(t) - K^2_{1/3}(t)\right]\right\}.
\end{eqnarray}
Here $t\equiv \nu/(3\gamma^2\nu_{\rm B})$ and $K_{\rm b}$ is the
modified Bessel function of order $b$.  The total monochromatic
synchrotron luminosity (in the comoving frame) is then found by
integrating $j(\gamma,\nu)$ over the particle distribution and the
source volume.

\subsection{Synchrotron self--absorption}

The synchrotron self absorption spectrum produced by a power law
energy particle distribution with a low energy cut--off
is $F(\nu)\propto \nu^2$ instead of the familiar $\nu^{5/2}$.
The type of distribution can result from the injection 
of particles above some minimum energy and slow cooling 
(with respect to $t_{\rm cross}$).
This case is therefore relevant to us.
Here we treat synchrotron self--absorption exactly, making use of the
concept of ``synchrotron cross section" $\sigma_{\rm s}(\gamma, \nu)$
given in Eq. (2.17) of Ghisellini \& Svensson (1991), which we 
report for completeness:
\begin{equation}
\sigma_{\rm s}(\gamma, \nu)\, =\, {8\sqrt{3} \pi^2 \over 15} \,
{e\over B}\, {t \over \gamma^5}\, \left[ K^2_{4/3}(t)-
K^2_{1/3}(t)\right],
\end{equation}
where $e$ is the electron charge and a tangled magnetic field and an
isotropic distribution of pitch angles are assumed.  
The synchrotron absorption optical depth is then given by
\begin{equation}
\tau^s_\nu \, =\, \Delta R^\prime \int \sigma_{\rm s}(\gamma, \nu)
N(\gamma)\, d\gamma.
\end{equation}
Note that in this formalism the optical depth can be determined also
when $N(\gamma$) has abrupt cut--offs.

The resulting monochromatic synchrotron luminosity is isotropic in the
comoving frame and is calculated according to the usual radiation
transfer equation:
\begin{equation}
L_{\rm s}(\nu) \, =\, 4\pi V \left[ \int j(\gamma, \nu)
N(\gamma)d\gamma\right] { 1-e^{-\tau_\nu^s} \over \tau_\nu^s}.
\end{equation}

\subsection{Inverse Compton emission}

We calculate the inverse Compton emission if it occurs in
the Thomson regime, while neglect the Klein--Nishina regime by
adopting a scattering cross section equal to the Thomson one for
$\gamma x <3/4$ and zero otherwise.  $x=h\nu /(m_{\rm e} c^2)$ is the
dimensionless energy of the incoming photon. The $\delta$--function
approximation for the single particle spectrum is adopted, i.e. it is
assumed that the particle emission is peaked at $x_{\rm c} =
(4/3)\gamma^2 x$.  With these simplifications the resulting emitted
monochromatic power is
\begin{equation}
L_{\rm c}(x_{\rm c}) \, =\, {\sigma_{\rm T} c V m_{\rm e}c^2 \over 2}
\int N(\gamma)\gamma n(x) dx; \quad \gamma = \left( {3 x_{\rm c} \over
4 x}\right) ^{1/2}
\label{eq:IC}
\end{equation}
where $n(x)$ is the density of seed photons.
Eq. (\ref{eq:IC}) is calculated for each order of Compton scattering.
To calculate $L_{\rm c, n}(x_{\rm c})$ of the 
Compton order $n$ we must then use the photon density produced
by the previous order, $n(x)_{\rm n-1}$.
For the first order, the seed photons are the sum of the locally
produced synchrotron and the externally produced photons.
The spectrum  of the external seed photons (in the comoving frame)
is assumed to have a blackbody shape peaking at frequency 
$x_0 = \Gamma x_{\rm line}$, where we always assume 
$\nu_{\rm line} =10^{15}$ Hz.  

As high energy $\gamma$--rays can interact with softer photons producing 
electron--positron pairs, we take into account the absorption effects
due to this process. However in general this is found to be
unimportant because no large compactenesses are involved.

\subsection{Beaming}

Finally let us consider the effect of relativistic beaming on the
observed radiation.  The emitting plasma is moving at a BLF
$\Gamma$, intermediate between the original BLF of the two
colliding shells, computed according to the details given in Section
2.  The radiation produced by each colliding shell is then received by
the observer as
\begin{equation}
L^{\rm obs}(\nu^{\rm obs}) \, =\, \delta^3 L(\nu^{\rm obs}/\delta), 
\end{equation}
where $\delta = [\Gamma -(\Gamma^2-1)^{1/2}\cos\theta]^{-1}$ is the
Doppler (or beaming) factor and $\theta$ is the viewing angle (with
respect to the jet axis).  The luminosity thus calculated corresponds
to what the observer would estimate assuming isotropic emission.

\section{Numerical simulations}
\label{sec:numsim}

In this section we analyze the general properties of the macro-- and
micro--physics of the plasma during the jet evolution.  Numerical
simulations are mandatory in order to test the model, given the
difficulty in understanding and disentangle the effect of each
parameter on the global properties of the flow and the produced
radiation.

The number of parameters that sets the kinematical and radiative
evolution of the jet flow is relatively small. More precisely, the
range of BLFs ($\Gamma_{\rm m} \div \Gamma_{\rm M}$), the
average kinetic luminosity of the inner engine $L_{\rm w}$ and the
time variability timescale $t_{\rm v}$ define the kinematical evolution,
while the equipartition parameters $\epsilon_{\rm e}$ and
$\epsilon_{\rm B}$ and the fraction of accelerated electrons 
$\zeta_{\rm e}$ control the radiative properties of the plasma.  
The former ones are reasonably constrained, since $10 \lsim \Gamma
\lsim 25$ and $L_{\rm w} \sim 10^{48}$~erg~s$^{-1}$ are
observationally inferred and the variability time can be
estimated as $t_{\rm v} \sim R_0/c \sim 10^4$~s.  
On the other hand, it is more difficult to quantify 
the latter ones and they can be basically considered as free 
parameters. 
In the simulations presented in this paper, they have been
specifically adjusted in order to reproduce the general observed
behavior of the OVV source 3C~279, as: $\epsilon_{\rm e} = 0.5$,
$\epsilon_{\rm B} = 0.004$ and $\zeta_{\rm e} = 0.04$.

Finally, the external radiation field (see above) is constrained by
the radius of the BLR, $R_{\rm BLR}=5\times 10^{17}$cm,
and by the luminosity of the disk and corona radiative components
assumed to be $L = 10^{45}$ erg s$^{-1}$.

In order to illustrate the difficulty in predicting the 
results of changes in the above parameters, let us consider, as an example, 
the effect of variations (decrease) in
the lower BLF $\Gamma_{\rm m}$.
On one side as the average speed of the shells is smaller, they must
be more massive to maintain the same wind luminosity.  This implies
that the energy per baryon generated in each collision is lower and
thus a smaller $\gamma_{\rm b}$ is expected.
On the contrary, the dynamical efficiency increases thanks to the
larger ratios of the BLFs of the involved shells, leading
in turn to larger internal energy, in direct competition with the
former effect.  Moreover, a smaller $\Gamma_{\rm m}$ causes the
collisions to start earlier, when the shell volume is smaller, with
the consequent creation/amplification of a larger magnetic field.
In summary, even a limited change in a single parameter causes several
(often competing) effects in the kinematical and
radiative evolution of the flow.  Therefore in the following we will
examine such consequences through numerical simulations of the
flow evolution.

\begin{figure}
\psfig{figure=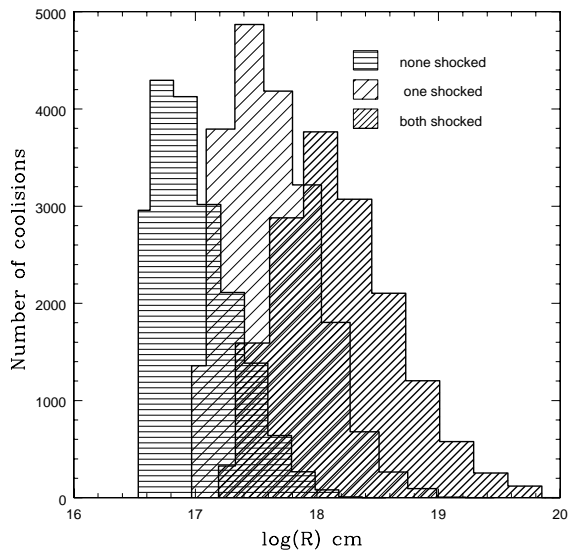,width=8cm}
\caption{{The histograms show
the distributions of radii of the collisions between shells that 
have: never collided before (i.e. with no internal energy), only one has  
collided before, both have collided before.} 
\label{fig:raggi}}
\end{figure}

\subsection{Output parameters}

A first key quantity calculated through the simulations is the
distribution of collision radii: this in fact controls the average
value of the magnetic field strength (which depends on the shell
volume, i.e. on the shock distance), and the number of collisions
occurring inside the BLR, which in turns determines the
relative importance of the radio--optical to $\gamma$--ray flux.

Figure~\ref{fig:raggi} shows the distribution of radii at which
collisions take place.
The limit at small distances directly follows from the choice of the
injection parameters, and its typical value can be approximated as:
\begin{equation}
R_{\rm min}\approx \frac{2 \alpha_\Gamma^2}{(\alpha_\Gamma^2-1)}
\Gamma_{\rm m}^2 c t_{\rm v} \sim 7 \times 10^{16}~~{\rm cm},~~~~~~
\alpha_\Gamma = \frac{\Gamma_{\rm M}}{\Gamma_{\rm m}}
\end{equation}
for an average time between two ejections $t_{\rm v}=10^4$s,
$\Gamma_{\rm M}=25$ and $\Gamma_{\rm m}=10$.
The merged shells collide again and again at larger distances up to
$\sim 10^{20}$ cm, where the differences among the shell velocities are
completely smoothed out and the wind can be considered uniform.

The overall duration of the simulation ($t_{\rm w}=10^8$s, corresponding to 
3$\times 10^4$ shells) has been chosen in order to reach 
a virtually steady state for the wind, i.e. with collisions 
taking place at all radii smaller than $\sim 10^{20}$~cm. 
As shown in Fig.~\ref{fig:raggi}, in the inner part ($R< 10^{18}$ cm)
the collisions are mostly between shells which have never collided
before, while shells colliding at larger radii
($10^{18}$--$10^{19}$cm) are the result of previous mergings. About 50
per cent of the shocks occur inside the BLR (for $R_{\rm
BLR} = 5 \times 10^{17}$~cm), while only 7.2 per cent take place at
radii $R > 10^{19}$~cm.

\begin{figure}
\psfig{figure=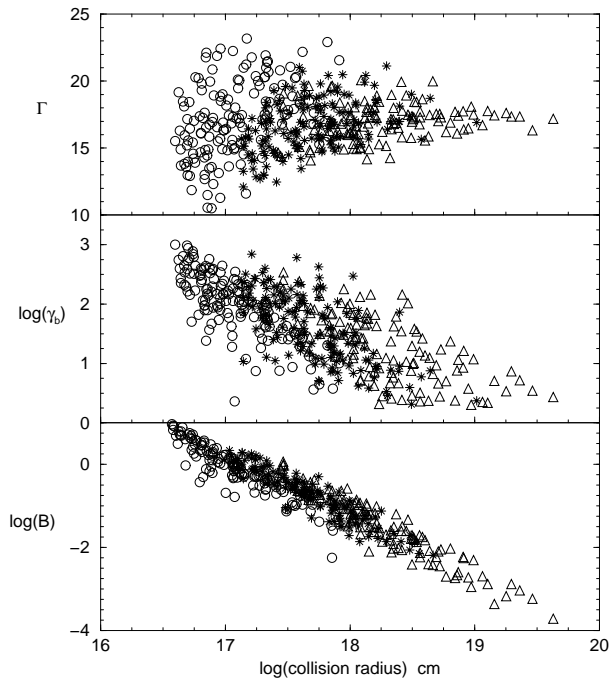,angle=-90,width=8cm}
\caption{Evolution of the post--shock shell parameters as a function
of the collision radius. The upper panel shows the BLF of the merged
shell after each collision; the central panel represents the minimum
RLF of the electrons $\gamma_{\rm b}$ (for $\epsilon_{\rm e} =
0.5$; $\zeta_{\rm e} = 0.04$); the lower panel shows the value of the
generated magnetic field ($\epsilon_{\rm B} = 0.004$). FS and
RS are considered independently.  Only one tenth of the points is
displayed for clarity.  Circles refer to collisions in which none of
the shells has ever collided before, pluses to collisions in which one
of the shells has collided, while triangles indicate collisions
between shells that have both collided before.
}
\label{fig:para}
\end{figure}

The net amount of internal energy $E_{\rm in}$ released in each
interaction depends on the dynamical efficiency and on the total kinetic
energy of the shells available before the collision.
As illustrated in the upper panel of Figure \ref{fig:para} the shell 
BLFs tend to average out during the wind expansion, leading 
to a strong attenuation of the dynamical efficiency: this can be as 
high as 10 per cent for the inner collisions but decreases to less 
then 0.1 per cent at the largest radii.
However, the outer shocks involve larger kinetic energies, as the
shells are much more massive -- being the result of many previous
collisions.
As a consequence, the average net amount of internal energy generated by
each interaction is roughly constant with radius, although as few
shells are left in the large scale wind, their rate of collisions is low.

As discussed in the previous section, the internal energy $E_{\rm in}$
is shared among protons, electrons and magnetic field.  
The central panel of Fig.~\ref{fig:para} shows the value of the minimum RLF
of the accelerated electrons, $\gamma_{\rm b}$, while the lower panel of
Fig.~\ref{fig:para} represents the magnetic field strength $B$ for the
same collisions. In some cases, the efficiency of the collision
is so small that the minimum Lorentz factor of the accelerated electrons
is lower than 2. Also the magnetic field is usually small in these cases.
These collisions are not considered in the radiative part of the code, 
since the cooling time of the electrons is so long that adiabatic losses
are overwhelming. In any case, the minimum Lorentz factor of the 
electrons always excedes the thermal motion of the fraction of
non accelerated ones.
Since $E_{\rm in}$ is approximately constant with radius, the value of 
$B$ strongly anti--correlates with the collision distance
approximately as $B \propto R^{-1.5}$, being inversely
proportional to the square root of the volume.

\begin{figure}
\psfig{figure=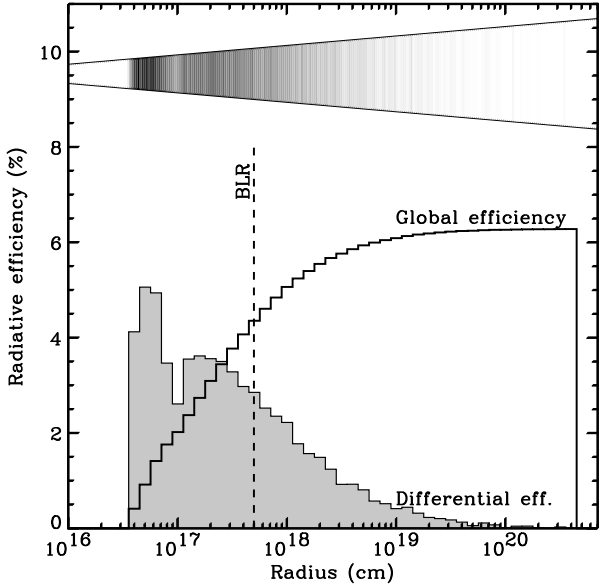,width=8cm}
\caption{The radiative efficiency versus the collision radius.  The
solid line refers to the global efficiency, i.e.  the fraction of the
total kinetic energy of the wind ($E_{\rm w}=L_{\rm w}\times t_{\rm w}$) 
radiated
on scales smaller than a given radius; the shaded hystogram shows instead
the differential efficiency, namely the fraction of $E_{\rm w}$
radiated for a given radius interval. The hystogram amplitude has been 
multiplied by a factor of 10 for clarity.  The vertical line indicates the
radius assumed for the extension of the BLR. The cone--like insert in the 
upper part of the figure shows a grey--tone representation of the 
differential efficiency of the jet. The darker the color the higher the 
efficiency (cfr. the shaded hystogram).}
\label{fig:effi}
\end{figure}

We can calculate what fraction of the energy produced at each location 
of the jet is used to accelerate electrons to relativistic energy.
In the absence of non--radiative losses, this energy will be 
eventually entirely radiated.
We then call this quantity, which is independent of the specific 
radiation mechanism and the timescales involved, the
``differential radiative efficiency".
The logarithmic differential efficiency, i.e. the efficiency per 
unit logarithmic radius, is shown in Fig.~\ref{fig:effi}, where we
can see that the most radiatively efficient region is the inner jet 
(between $10^{16.5}$ and $10^{18}$~cm), where the collision 
rate is the highest. 

The corresponding integrated efficiency, i.e. the fraction 
of $L_{\rm w}$ radiated from the beginning of the jet up to a 
given radius, is shown in Fig.~\ref{fig:effi}: it saturates at 
$R \sim 10^{19}$ cm at a value of $\sim$ 6 per cent.
Such value can in principle depend on the input parameters,
but an efficiency of a few per cent is a robust result unless a very broad
distribution of the initial BLFs is involved \cite{B00}.
Note that the finite duration of the computation 
does not affect the efficiency estimate, since the differential value
approaches zero at radii $\sim 10^{19}$ cm, much smaller than 
the maximum radius reached in the simulation. 

\begin{figure}
\psfig{figure=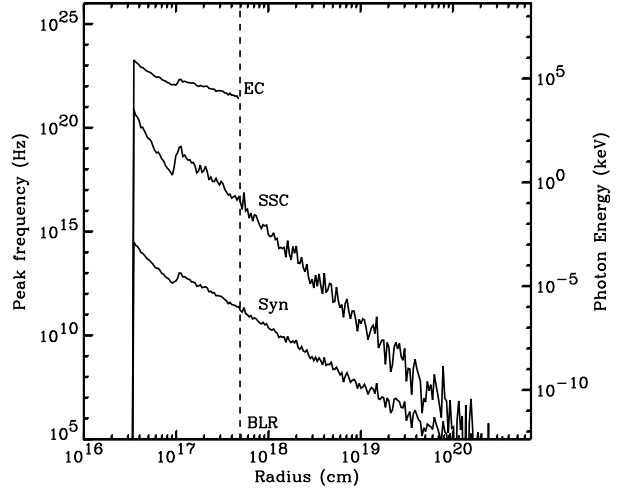,width=8cm}
\caption{Evolution of the peak frequency of radiation pulses following
collisions occurring at different radii, for synchrotron (bottom line),
SSC (median line), and EC (upper line).}
\label{fig:pic}
\end{figure}

From the micro--physical properties of the plasma shown 
in Fig.~\ref{fig:para} it is also possible to 
compute the typical frequencies of the emitted photons.
For the inner collisions ($R < R_{\rm BLR}$) synchrotron, SSC and 
EC are responsible for the observed radiation. 
At larger radii, external to the photon bath provided by the BLR 
re-processing, the first two mechanisms dominate the emission.
The evolution of the peak frequencies corresponding to
the three radiative processes along the jet  is shown in 
Fig.~\ref{fig:pic}, where
their values at each radius have been obtained by averaging over all
of the collisions occurring at that radius.  The synchrotron peak
frequency $\nu_{\rm syn}\propto \gamma_{\rm b}^2 B$ decreases from
$\sim 10^{15}$ Hz for the initial collisions to $\sim$ 1 GHz at pc
scale. The SSC peak frequency, given by $\nu_{\rm
SSC}\propto\gamma_{\rm b}^2 \nu_{\rm syn}$, varies from $\sim 10^{20}$
to $\sim 10^{16}$ Hz, for $R$ increasing from $3\times10^{16}$ to
$10^{18}$ cm.  Finally the EC typical frequency $\nu_{\rm EC}\approx
\Gamma_{\rm m}^2\gamma_{\rm b}^2 \nu_{\rm line}\approx 10^{23}$Hz
(for $\nu_{\rm line}=10^{15}$~Hz).

\begin{figure}
\psfig{figure=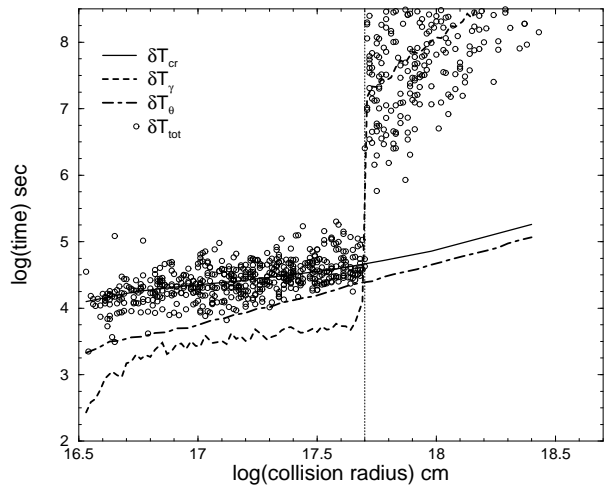,angle=-90,width=8cm}
\caption{Time scales determining the duration of the observed pulses
versus the collision radius. The dashed line refers to the angular
spread time, the solid line to the time a shock takes to cross the
shells, and the dotted line to the electron cooling time.  The circles
represent the pulse duration (for clarity we plot only one tenth of
the efficient collisions).}
\label{fig:time}
\end{figure}

Let us now consider the evolution of the effective radiative
efficiency, which is estimated by comparing the radiative time scales
$\delta t_{\gamma}$ of relativistic electrons (corresponding to the
cooling due to the three above processes) with the time a shock
takes to cross the shell width (i.e. the energy injection time)
$\delta t_{\rm cr}$, and the angular
spread time, $\delta t_{\theta}$ (see Fig.~\ref{fig:time}).
$\delta t_{\rm cr}$ increases from a few hr at $5\times 10^{16}$ cm to
$\sim$~a week at $10^{18}$ cm, because of both the increase in the
shell width and the slowing down of the shock velocity during the
expansion.  The angular spread time reflects the different distances
traveled by photons propagating along different directions. Since the
observed radiation is beamed in a solid angle $\sim 1/\Gamma$, $\delta
t_{\theta}$ corresponds to the delay in the arrival time of photons
emitted at an angle $1/\Gamma$ with respect to those propagating along
the line of sight: $\delta t_{\theta}\approx R/2c\Gamma^2$.  
This varies from a few hr to $\sim$~a week, analogously to the shell
crossing time--scale.  
The cooling time $\delta t_{\gamma}$ is
determined by the most efficient process among synchrotron, SSC and EC. 
Inside the BLR the latter dominates, with a cooling time of less
than a hr, over the $\sim$ 1 day time scale of synchrotron and SSC
emission. 
Outside the BLR, the synchrotron process dominates the
emission, with strongly radial dependent cooling times $t_{\rm syn}
\approx$ 1 yr 
\footnote{Clearly the synchrotron and SSC cooling
times increase with radius, since $\gamma_{\rm b}$ and $B$ decrease
during the expansion.}.
Note that inside the BLR the cooling is so efficient that electrons
emit most of their energy immediately after being accelerated and the
pulse duration is due to a combination of the shock crossing time and
geometrical effects.  
At these radii, the radiation is mostly emitted (via EC) at 
high frequencies, $\sim 10^{22}-10^{24}$Hz.  
Outside the BLR instead electrons cool on a time scale much longer 
than $\delta t_{\rm cr}$, and thus the pulse duration is set by the 
synchrotron cooling time and the peak frequency is in the mm range. 

\section{Results}
\label{sec:res}

Via the numerical simulations the full time--resolved spectral
behavior of the source can be determined.  In this section, we
directly perform comparisons of the model output quantities with
observational data relative to 3C~279, by decomposing the information
into (snapshots or time averaged) spectra and light curves.  The full
time--dependent behavior of the simulated source can be also examined
through animations (see {\it
http://www.merate.mi.astro.it/$\sim$lazzati/3C279/}) in which the
temporal and spectral evolutions are simultaneously shown.

\begin{figure}
\psfig{figure=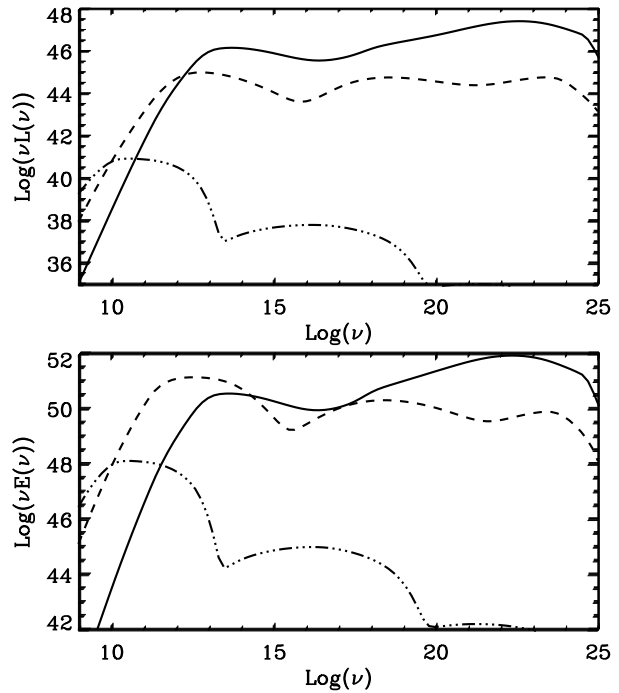,width=8cm}
\caption{Average spectra of collisions. Solid line show averages
for collisions happening inside the broad line region $R < 5\times 
10^{17}$~cm, dashed line shows collisions happening at intermediate 
distances $ 5\times10^{17} < R < 10^{19}$~cm and dot--dashed line
show collisions at large radii ($R>10^{19}$~cm). Upper panel show
the average spectrum of the shells without taking into account the 
duration of the emission. The Figure show the average spectrum of a 
single shell in the given range of radii. In the lower panel,
instead, the duration of the emission of each collision has been
taken into account. 
The plot shows the average instant spectrum of the 
whole section of the jet.}
\label{fig:spettri}
\end{figure}

The first step of the simulations (see the previous section)
determines a set of physical parameters for every collision.  In a
second step these are then used to produce the Spectral Energy
Distributions (SED) emitted in each collision (one SED for the
forward and an independent SED for the reverse shock) as
described in Section~\ref{sec:spec}.
Different spectra are generated in different collisions, but general
behaviors which  describe the overall shape of the SED as a
function of radius 
can be found. As already stressed the most important condition
that affects the SED shape is whether the collision occurs inside or
outside the BLR.  In Figure~\ref{fig:spettri} we show the averaged
spectra of collisions taking place within a given range of radii. Shocks
inside the BLR produce spectra dominated by the EC process, through
the scattering of electrons with energy $\gamma_{\rm b} m_{\rm e}c^2$
on BLR photons.  
The synchrotron peak, due to the same electrons, is located in the
near infrared.  A minimum occurs in the optical--UV range, while the
X--ray band is characterized by a power--law spectrum.  Outside the
BLR (lacking the EC component) the SED becomes more and more dominated
by the synchrotron process: immediately outside the BLR the Compton
parameter $y \sim$ unity and the spectrum is the sum of the
synchrotron and the first two Compton orders; as the radius increases,
both the relativistic particle density and their average (squared)
energy decrease, leading to progressively smaller values of $y$ (if
only radiative losses are considered). At the same time the
synchrotron self--absorption frequency decreases with radius.

\subsection{The observed spectrum}

We have so far examined the role of the various emission mechanisms at
different distances along the jet.  The observed spectrum is the
convolution of the emission of all photons arriving at the observer
simultaneously: since these are produced at different distances, it is
necessary to take into account the different light propagations times
when summing them up. In other words, the resulting spectrum is the 
sum of all the spectra produced by shocks 
simultaneously active in the observer frame, taking into
account the radius at which each collision is occurring and the
travel path of photons to reach the observer.

We therefore associate to each spectrum a photon pulse, with a start
time (the time at which the collision begins) and a duration (set by
the combination of the timescales discussed in
Section~\ref{sec:numsim}). 
The shape of such pulse is triangular if its duration is determined  
by geometrical effects and/or by the shock crossing time  
(with a FWHM equal to the timescale shown in Fig.~\ref{fig:time}): 
these are generally associated to collisions inside the BLR. 
If the pulse duration is instead set by cooling the pulse rises
linearly, as in the previous case, but after the maximum it decays
exponentially [$\propto \exp(t/3 t_{\rm cool}(\gamma_{\rm b}) )$]: 
in this case the rise and decay times can be very different.

\begin{figure}
\psfig{figure=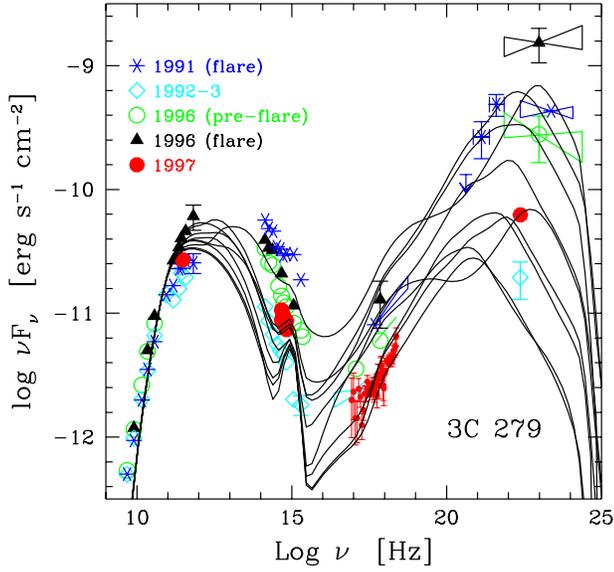,width=9cm}
\vskip -1 true cm
\caption{Example of the spectra predicted by our model compared
with the SED of 3C 279 corresponding to different observational campaigns.}
\label{fig:spectras}
\end{figure}

\begin{figure}
\psfig{figure=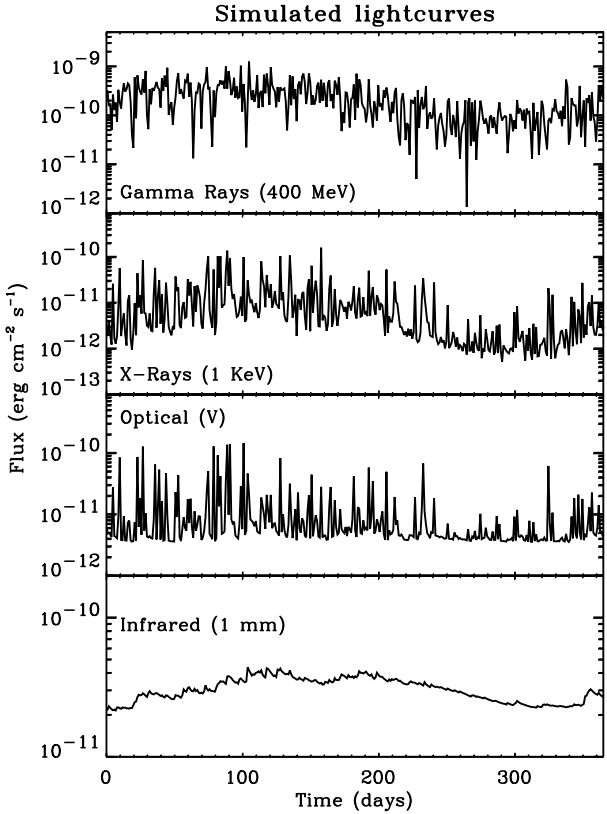,width=8cm}
\caption{Light curves at different frequencies. The $\gamma$-rays,
$X$-rays and Optical light curves vary on different time scale, with a
minimum value of few hr, and the IR one varies on few months time
scale.}
\label{fig:lcur}
\end{figure}

\begin{figure}[!t]
\psfig{figure=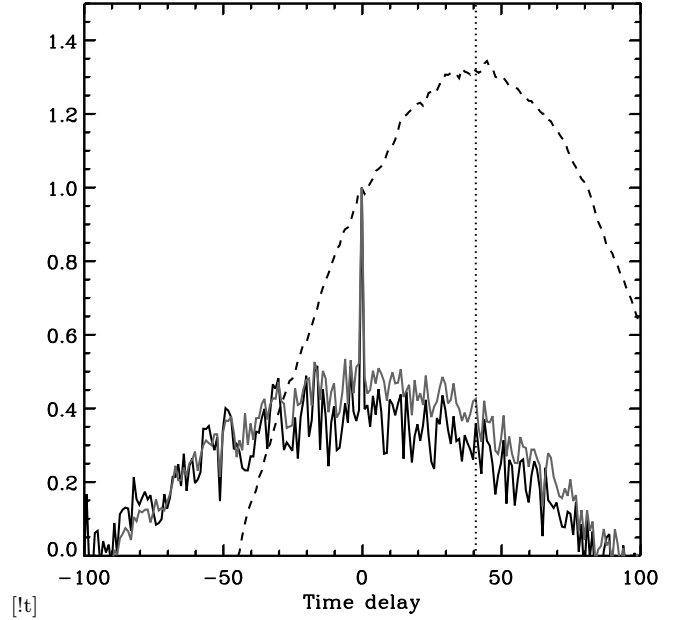,width=8cm}
\caption{Results of cross correlations between $\gamma$-rays and
X--ray (solid grey line), optical (solid dark line) and infrared (dashed 
line) light curves. The X-rays and optical emission vary simultaneously 
with the $\gamma$-rays, while the IR follows the $\gamma$-rays emission with a
delay of a $\sim 40$~days (the vertical dotted line).}
\label{fig:cross}
\end{figure}

\subsection{The spectrum of 3C 279}

Examples of the resulting spectra are compared to data from three
different simultaneous multiwavelength campaigns of 3C279 in
Figure~\ref{fig:spectras}.
Note the remarkable variability, especially at $\gamma$--ray energies.

Within the model examined here the source high state corresponds to a time
interval in which at least one of the active regions is inside the BLR.  The
lower line shows the SED in a quiescent state of the source, i.e. when
all of the active collisions are located outside the BLR.  The difference
between the two spectra manifests itself mainly in the high frequency
range, where the internal collisions, that have a shorter pulse
duration, mostly contribute.  
Although the number of external collisions 
is almost equal to the number of the inner ones, 
the former produce longer pulses: it follows that the number of those 
simultaneously active is higher and hence the statistical variability is 
suppressed.  

A different way to examine the results is to determine light
curves in given energy bands: Figure~\ref{fig:lcur} shows the flux
temporal behavior in the $\gamma$--ray, X--ray, optical and millimeter
ranges.  The high frequency fluxes (down to optical energies)
are highly variable (again since they are produced mainly by the inner
collisions), while the millimeter flux comes from outer
shocks and thus the corresponding light curve is much smoother.

In order to study the presence of delays in the variability at
different wavelengths, we added ``secular'' variations in the kinetic
luminosity injected by the inner engine. In particular, this is
modulated as a sinusoidal function with period $\sim$ 1 year, spanning
a factor of five in luminosity (between the minimum and the maximum).
Figure~\ref{fig:cross} illustrates the results of the cross correlations
between the $\gamma$--ray and the other frequencies light curves.
Once again the plot shows that $\gamma$--ray, X--ray and optical photons are
produced in the same region (being closely correlated, with no time delay),
while the IR emission lags by  $\sim 40$~days, 
as the shells that produce high frequency photons inside
the BLR have to travel to larger radii ($R \sim 10^{19}$~cm) to produce
IR emission. The expected observed delay is given by:
\begin{equation}
\Delta t = {{\Delta R} \over {c \Gamma^2}} \sim 38.5 \;
\Delta R_{19} \; \Gamma_1^{-2} \;\; {\rm days}, 
\end{equation}
which indeed well describes the results of the simulations.

Finally note that time delays of the order of $\sim$ hr might occur
between $\gamma$--ray and X--ray/optical peaks even if they are
produced in the same region. 
However, the study of such delays would require to take into
account the different light travel paths within a single region
(Chiaberge \& Ghisellini 1999), and a time dependent treatment 
of the shell emission, taking into account the variations in the 
magnetic field and electron distribution during the cooling.  
Such a treatment is beyond the scope of the present work and will be 
examined in future developments of the model.

\section{Discussion}

Highly relativistic flows dominate the dynamics and emission in both
Gamma--Ray Bursts and blazars. It has been proposed that internal
shocks, produced by the interaction of flow components injected with
different bulk speeds, can be responsible for the observed emission
and variability in both classes of sources (Rees 1978; Rees \&
Meszaros 1994).

Here we have considered in some details this `internal shock' scenario
in the context of AGN jets, self--consistently accounting for the
dynamics and the radiative properties of the flow. After describing
the kinematic and radiative assumptions of the model, we have examined
the results of numerical simulations which allow us to study the
predicted energy deposition on different jet scales, the relevant
timescales, the corresponding dominating emission processes, the
spectral evolution and predicted variability, the correlations among
the emission at different frequencies.

The key interesting features of this scenario are the relatively low
radiative efficiency (of order of a few per cent) which well accounts
for the dissipation of kinetic energy in blazars, as higher
dissipation rates would create major energetic problems (in order to
carry sufficient power up to the large scale lobe
structures). Furthermore, although the radiative dissipation occurs on
all jet scales, as indeed observed, the bulk of it is localized at
tenths of a pc, on the BLR scale, in agreement with the requirements
of fast variability and transparency to $\gamma$--rays.

The results of the simulations, based on input parameters suitable to
qualitatively reproduce the observations of a powerful radio--loud
quasar, such as 3C 279, are indeed satisfactory.  As expected the
plasma in the inner part of the jet cools predominantly via EC
emission, while on larger scales synchrotron cooling dominates.
Qualitative agreement is also obtained in terms of light curves, and
correlations among different spectral bands are predicted. Also, an
asymptotic value of the bulk Lorentz factor $\Gamma \sim 15$ is
achieved at distances $\sim$ pc.

While this paper has been aimed at presenting the model and study its
general applicability to blazar jets, the final goal is to determine
which initial parameters/conditions are required in order to produce
the whole of the blazar phenomenology and energetics. In particular,
the model significantly constrains the typical Lorentz factors (and
their distribution) achieved through acceleration on the inner scales,
and the typical timescale of plasma injection. 
In this respect, also the determination of the duty cycle of
the $\gamma$--ray emission in blazars is of chief importance, as it
can be used to constrain the rate and Lorentz factor distribution of
the injection of energy/plasma in the relativistic flow. Note that
although more accurate measures of $\gamma$--ray spectra and the
detection of a much larger number of $\gamma$--ray loud sources will
require the capabilities of GLAST, tight constraints on the duty
cycles will be available much sooner thanks to the launch of
AGILE. The model presented here will provide (for the first time) the
possibility of directly modelling the time dependent broad band
spectral evolution of these sources following the (time dependent)
energy injection.  The relationship between flares at high energies
and the birth of superluminal radio knots and/or radio flares will be
also explored.

Therefore issues associated with the jet energy content and the
relative conversion of bulk energy into internal (particle and field)
energy at relativistic shocks can be also constrained by the
comparison of the model predictions with observations.

A further aspect we intend to examine is the propagation and
dissipation of jets on larger scales, where however uncertainties are
large. 
On the $\gg$ pc--scales, significant entrainment might occur,
possibly leading to the formation of external shocks and the
development of slower outer layers around jets. 
Another issues is the relative importance of creation vs amplification of an
existing seed magnetic field: if the latter process prevails, it will 
probably lead to larger magnetic fields on the kpc scale jet, and
consequently to faster synchrotron cooling.
Inverse Compton cooling at kpc scales is instead enhanced, with 
respect to what calculated here, by the interaction of with the 
cosmic background radiation, which is seen relativistically boosted 
in the rest frame of the fast moving parts of the jets
(e.g.  Celotti, Ghisellini \& Chiaberge, 2000).

We finally stress that it is of primary importance, for the
understanding of the flow injection and its connection with the
central accreting engine, to determine the key parameters which
regulate the observed spectral and luminosity trend, from high power,
low energy peaked, flat spectrum radio quasars, to weak high energy
peaked BL Lacs (Fossati et al. 1998). This will be the content of a
forthcoming paper.

\section{Acknowledgments}
We wish to thank the anonymous referee for his/her 
constructive and careful comments.
The Italian MURST is acknowledged for financial support (AC).

\end{document}